\def \SAIT #1 #2 {{\em Mem.\ Soc.\ Astron.\ It.\/} {\bf #1}, #2}
\def \MESS #1 #2 {{\em The Messenger\/} {\bf #1}, #2}
\def \ASTRNACH #1 #2 {{\em Astron. Nach.\/} {\bf #1}, #2}
\def \AAP #1 #2 {{\em Astron. Astrophys.\/} {\bf #1}, #2}
\def \AAL #1 #2 {{\em Astron. Astrophys. Lett.\/} {\bf #1}, L#2}
\def \AAR #1 #2 {{\em Astron. Astrophys. Rev.\/} {\bf #1}, #2}
\def \AAS #1 #2 {{\em Astron. Astrophys. Suppl. Ser.\/} {\bf #1}, #2}
\def \AJ #1 #2 {{\em Astron. J.\/} {\bf #1}, #2}
\def \ANNREV #1 #2 {{\em Ann. Rev. Astron. Astrophys.\/} {\bf #1}, #2}
\def \APJ #1 #2 {{\em Astrophys. J.\/} {\bf #1}, #2}
\def \APJL #1 #2 {{\em Astrophys. J. Lett.\/} {\bf #1}, L#2}
\def \APJS #1 #2 {{\em Astrophys. J. Suppl.\/} {\bf #1}, #2}
\def \APSS #1 #2 {{\em Astrophys. Space Sci.\/} {\bf #1}, #2}
\def \ASR #1 #2 {{\em Adv. Space Res.\/} {\bf #1}, #2}
\def \BAIC #1 #2 {{\em Bull. Astron. Inst. Czechosl.\/} {\bf #1}, #2}
\def \JSQRT #1 #2 {{\em J. Quant. Spectrosc. Radiat. Transfer\/} {\bf #1}, #2}
\def \MN #1 #2 {{\em Mon. Not. R. Astr. Soc.\/} {\bf #1}, #2}
\def \MEM #1 #2 {{\em Mem. R. Astr. Soc.\/} {\bf #1}, #2}
\def \PLR #1 #2 {{\em Phys. Lett. Rev.\/} {\bf #1}, #2}
\def \PASJ #1 #2 {{\em Publ. Astron. Soc. Japan\/} {\bf #1}, #2}
\def \PASP #1 #2 {{\em Publ. Astr. Soc. Pacific\/} {\bf #1}, #2}
\def \NAT #1 #2 {{\em Nature\/} {\bf #1}, #2}

\documentstyle{memsait}
\input epsf.sty
\begin{opening}
\title{{\it BEPPO }SAX OBSERVATIONS OF PKS 2155-304 DURING AN ACTIVE GAMMA-RAY STATE}
\author{MARASCHI L.$^2$, CHIAPPETTI L.$^1$, TAVECCHiO F.$^2$ CELOTTI A.$^3$, FOSSATI
G.$^3$ GHISELLINI G.$^2$, GIOMMI P.$^4$, PIAN E.$^5$, TAGLIAFERRI G.$^2$,
TREVES A.$^6$, ZHANG Y.H.$^3$}
\institute{$^1$ IFCTR-CNR, Via Bassini 15 Milano\\
$^2$Osservatorio Astronomico di Brera, Via Brera 28, Milano\\
$^3$SISSA/ISAS, Via Beirut 2-4, Trieste\\
$^4$Beppo-SAX SDC, A.S.I., Via Corcolle 19, Roma\\
$^5$ITESRE-CNR, Via Gobetti 101, Bologna\\
$^6$Dipartimento di Fisica, Universit\'a di Milano sede di Como, Via Lucini 3, Como}
\date{} 
\end{opening}

\begin{document}

\oddpagefooter{}{}{} 
\evenpagefooter{}{}{} 
\ 
\bigskip

\begin{abstract}
PKS 2155-304 was observed with BeppoSAX in November 1997 for 64 ksec 
(total elapsed time 33.5 hours) and, for the first time,
 simultaneously in $\gamma$-rays with EGRET on
board the Compton Gamma Ray Observatory and with the ground based TeV
telescope CANGAROO, during a phase of high brightness in the X-ray band.
The LECS and MECS light curves show a pronounced flare (with an excursion of
a factor 3.5 between min and max), with
evidence of spectral hardening at maximum intensity.
The source is weakly detected by the PDS in the 12-100 keV band with no
significant evidence of variability. 
The broad band X-ray data from Beppo SAX are compared with the gamma-ray
results and discussed in the framework of homogenous synchrotron self Compton
models. 
\end{abstract}

\section{Introduction}

PKS 2155-304 is one of the brightest BL Lacertae objects in the X-ray
band and one of the few detected in $\gamma$
-rays by the EGRET experiment on  CGRO (Vestrand et al. 1995).
No observations at other wavelengths simultaneous with the gamma-ray ones
were ever obtained for this source, yet it is essential to measure the
Compton and synchrotron peaks at the same time in order to constrain
emission models unambiguously (e.g Dermer et al. 1997, Tavecchio et al. 1998a,b).
For these reasons having been informed by the EGRET team of their
observing plan and of the positive results of the first days of the CGRO
observation, we asked to swap a prescheduled target of our $BeppoSAX$
blazar program with PKS 2155-304. During  November 11-17 1997 (Sreekumar
\& Vestrand 1997) the $\gamma$
-ray flux from PKS 2155-304 was very high, roughly a factor of three
greater than the previously published value from this object. $BeppoSAX$
pointed PKS 2155-304 for about 1.5 days starting Nov 22. 
Here we report and discuss the data obtained from the $BeppoSAX$
observation. A complete paper including a detailed description of the data analysis
procedure is in preparation (Chiappetti et al. 1998) and we plan to submit it jointly
with a full EGRET paper (Vestrand et al. 1998).

\section{Results}

\subsection{Light curves}
Here we
summarize some of the results.
Fig 1 (left frame)  shows the light curves  binned over 1000 sec obtained
 in different energy bands.
 The light curves show a clear high amplitude
variability:  three peaks can be identified.  
The most rapid variation observed (the decline from the peak
at the start of the observation) has a halving timescale 
of about $2\times  10^{4}$ s, similar to previous occasions
(see e.g. Urry et al. 1997). No shorter time scale variability is detected although we
would have been sensitive to doubling timescales of order $10^{3}$ s.
 The variability amplitude is energy dependent as shown by the hardness ratio
histories plotted at the bottom of Fig 1.
The  HR  correlates positively with the flux, indicating that states with
higher flux have harder spectra.

\begin{figure}
\epsfysize=9cm 
\hspace{-0.2cm}\epsfbox{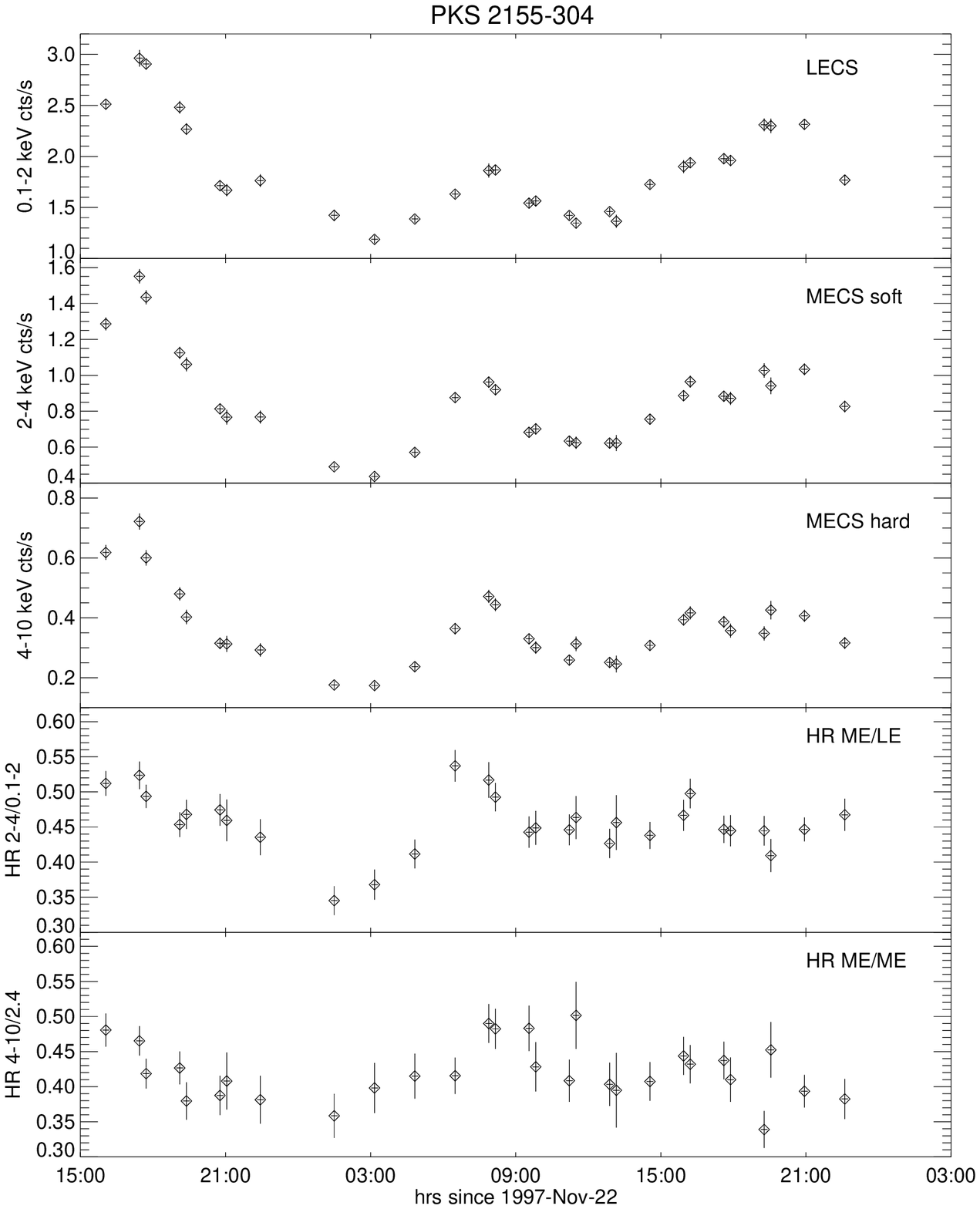}
\epsfysize=8.5cm 
\vspace{-8.75cm}
\hspace*{5.7cm}\epsfbox{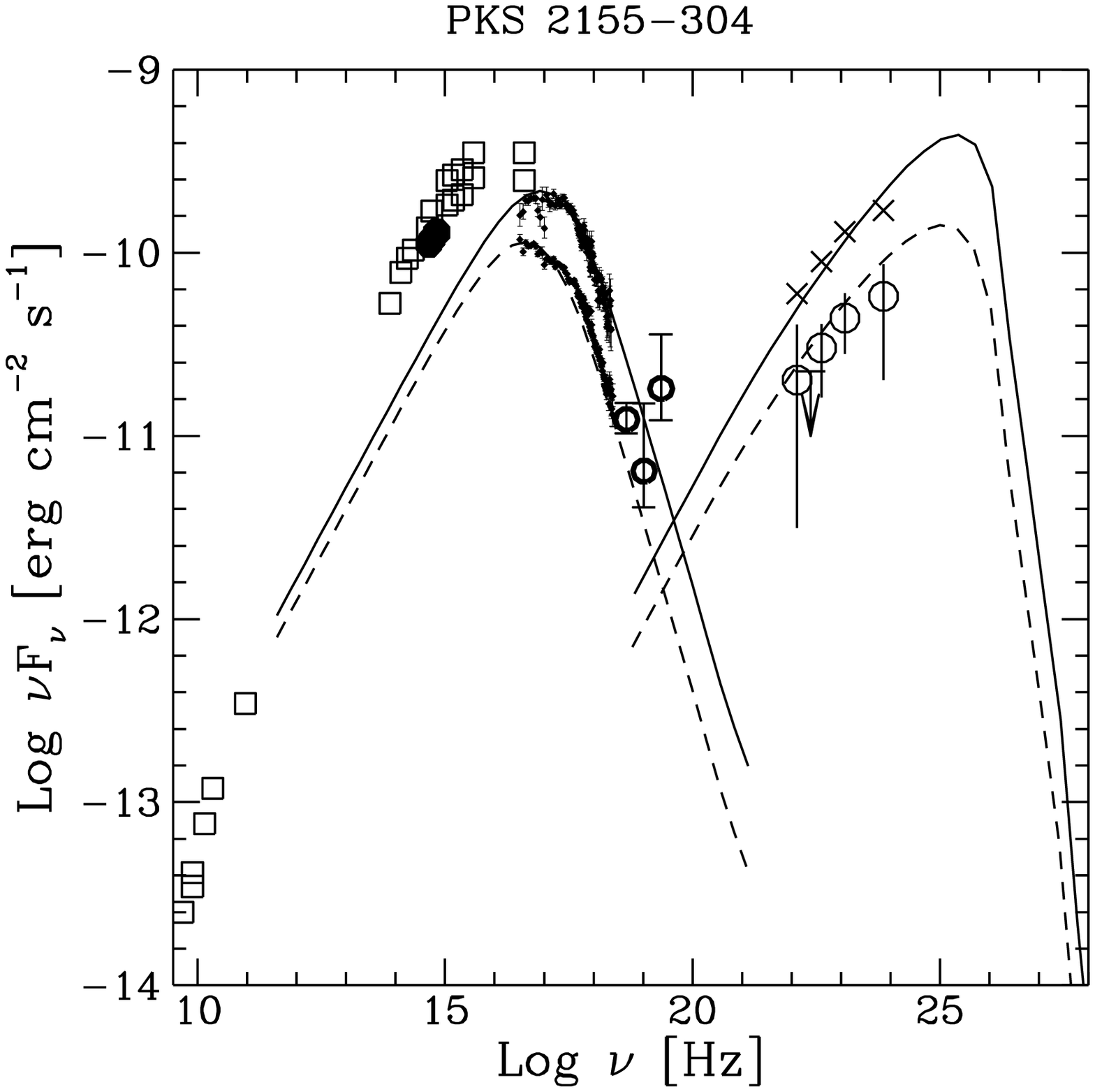}
\caption[h]{Left frame: The first three panels show the light curves in the
0.1-2 keV  (LECS) and 2-4 keV and 4-10 keV (MECS) energy bands respectively.
The last two panels show the 2-4 keV /0.1-2 keV and
the 4-10 keV/2-4 keV hardness ratios. Right frame: SED of PKS 2155-304
with the spectra 
calculated with the SSC model. The value of the parameters (see text) are: $\delta =18$, $B$=1 G, $K=
10^{4.68}$, $R=3\cdot 10^{15}$ cm, $\gamma _b=10^{4.49}$, $n _1=2$, $n_2=4.85$ 
(low state, dashed line)
and $\delta =18$, $B$=1 G, $K=
10^{4.8}$, $R=3\cdot 10^{15}$ cm, $\gamma _b=10^{4.65}$, $n _1=2$, $n_2=4.85$ 
(high state, solid line) 
}
\end{figure}

\subsection{Spectral analysis}

We found that the LECS and MECS spectra are individually well fitted
by a broken power law model with galactic absorption ($N_H=1.36 \cdot 10^{20}
cm^{-2}$, while single power law fits are
unacceptable. 
The fitted spectral parameters are
given in Table 1. The change in slope between the softest (0.1-1 keV) and hardest
 (3-10 keV) bands is $\simeq 0.8$ A broken power law fit to the combined
LECS and MECS spectra yields unsatisfactory results indicating that the 
spectrum has a continuous curvature. 

Fitting together the  MECS and PDS data yields  spectral parameters 
very similar to those  obtained for the MECS alone. The residuals show
that the PDS data are
consistent with an extrapolation of the MECS fits up to about 50 KeV.
Above this energy the PDS data show indication of an excess, indicating a
flattening of the spectrum.

\vspace{1cm} 
\centerline{\bf Tab. 1 - Spectral parameters}

\begin{table}[h]
\hspace*{0.8cm} 

\begin{tabular}{|c|c|c|c|c|c|c|} \hline
Data Set &  \multicolumn{2}{c|}{Power-law} & \multicolumn{4}{|c|}{Broken Power-law}\\ \hline \hline
 & $\Gamma$ & $\chi ^2_r$ & $\Gamma _1$ & $\Gamma _2$ & $E_{break}$ & $\chi ^2_r$ \\ \hline \hline
LECS 0.1-4 keV&  $2.21 \pm 0.01 $& 10.8 & $2.06 \pm 0.02$ & $2.54 \pm 0.04$ & $ 1.1 \pm 0.7$ & 1.221 \\ 
MECS 2-10 keV&  $2.78 \pm 0.02$  & 2.1 & $2.64_{-0.08}^{+0.06}$ & $2.88_{-0.03}^{+0.07}$ & $3.2\pm 0.8$ & 1.384 \\ \hline
\end {tabular}
\end{table}

\section{Spectral Energy Distributions and Discussion}

The deconvolved spectral energy distributions (SED) measured by SAX (0.1-300 keV)
at maximum and minimum intensity during this observation are compared in Fig.
1 (right frame) with non simultaneous data at lower frequencies and
 with the gamma-ray data
from the discovery observation (Vestrand, Stacy and Sreekumar 1995). The latter 
are also shown multiplied by
a factor three to represent the gamma-ray state of November 1997 as communicated in
IAU circular (Sreekumar \& Vestrand 1997). The final $\gamma$-ray data are not
available yet. From the public X-ray data obtained by the All Sky Monitor on XTE we
infer that the source was brighter during the first week of the EGRET pointing,
 which yielded the high $\gamma$-ray flux (Sreekumar \& Vestrand 1997 ) than during 
the {\it Beppo }SAX
observations. We therefore suppose that the $\gamma$-ray flux simultaneous to our
observations could be intermediate between the two states reported in the figure.
Note also that the PDS data refer to an "average" 
state over the SAX exposure time.

In order to estimate the physical parameters of the emitting region
in PKS 2155-304 we fitted the observed SEDs in the full X-ray range
with a simple  SSC model
involving  a homogeneous spherical region of radius R, magnetic field B, filled
with relativistic particles with energy distribution described by a broken power law
(4 parameters: $n_1$, $n_2$, $\gamma _b$ and a normalization constant, $K$) and with Doppler
factor $\delta$. This seven parameter model is strongly constrained by the data
which yield a determination of the two slopes (X-ray and gamma-ray slope) the 
frequency and flux of the synchrotron peak, a flux value for the Compton
component and a lower limit to the Compton peak frequency.  
Assuming $R=ct_{var}$ with $t_{var}=2$ hours the system is practically closed. 
 A general discussion of the parameter determination procedure for this class 
of models, with analytic formulae is given in Tavecchio et al. 
(1998a, b). 

In Fig.1 we show two models representing the high and low X-ray
intensity intervals in our observation.
We arbitrarily assumed that the lower intensity state corresponds to the gamma-ray
intensity reported in 1995 and we chose not to fit the low frequency data since there
the variability time scales are longer and they could refer to a larger emission region.
In order to account for the flaring state the break energy of the electron
spectrum was shifted to higher energies, leaving the other 
parameters unchanged.
Correspondingly also the Compton peak increases in flux and shifts to higher energies.
Both effects are however reduced with respect to the "quadratic" relation expected
in the Thomson limit, since for these very high energy electrons  the Klein-Nishina
limit plays an important role.
The predicted TeV flux is  $F(>1 TeV)= 10^{-11}$ ph cm$^{-2}$ s$^{-1}$ and 
 $F(>1 TeV) =2.5 \cdot 10^{-12}$ ph cm$^{-2}$ s$^{-1}$ in the two states respectively. 
Unfortunately
the CANGAROO telescope did not detect PKS 2155-304 in November 1997, but the upper
limits are consistent with the predicted values (Kifune, priv. comm.).
The sensitivity of the CANGAROO observatory is expected to improve significantly
in the next year, with the addition of new telescopes. It will therefore be
worthwhile to repeat the "experiment" of simultaneous X-ray and TeV observations
to verify whether the predicted TeV flux is actually observed.

\end{document}